\documentclass{aa}
\usepackage{amsmath,graphicx,amssymb}
\usepackage{txfonts}
\usepackage{natbib}
\bibpunct{(}{)}{;}{a}{}{,}

\newcommand\hvezda{\object{CU\,Vir}}
\newcommand{\zav}[1]{\left(#1\right)}
\newcommand{\hzav}[1]{\left[#1\right]}
\newcommand{\szav}[1]{\left\{#1\right\}}
\newcommand\intvidpo{\!\!\int\limits_{\begin{array}{c}\text{\scriptsize
visible}\\[-2mm]\text{\scriptsize surface}\end{array}}\!\!}

\newlength\staretab
\newcommand{\ms}{\ensuremath{{M}_{\odot}}}
\newcommand{\msr}{\ensuremath{\ms\,\text{yr}^{-1}}}
\newcommand\kms{\ensuremath{\text{km}\,\text{s}^{-1}}}
\newcommand\de{\text{d}}

\makeatletter
\def\sgn{\mathop{\operator@font sgn}\nolimits}
\makeatother
\def\clap#1{\hbox to 0pt{\hss#1\hss}}
\renewcommand\dot[1]{\overset{\mathrm{\raisebox{-0.15ex}{\clap{$\displaystyle\hspace{0.1em}.$}}}}{#1}}
\renewcommand\ddot[1]{\overset{\mathrm{\raisebox{-0.15ex}{\clap{$\displaystyle\hspace{0.1em}.\hspace{-0.08em}.$}}}}{#1}}
\renewcommand\dddot[1]{\overset{\mathrm{\raisebox{-0.15ex}{\clap{$\displaystyle\hspace{0.1em}.\hspace{-0.1em}.\hspace{-0.1em}.$}}}}{#1}}

\defcitealias{mycuvir}{Paper~I}

\begin{document}

\title{HST/STIS analysis of the first main sequence pulsar CU\,Vir\thanks{Based
on observations made with the NASA/ESA Hubble Space Telescope, obtained
at the Space Telescope Science Institute, which is operated by
the Association of Universities for Research in Astronomy, Inc., under NASA
contract NAS 5-26555. These observations are associated with program \#14737.}}

\author{J.~Krti\v{c}ka\inst{1} \and Z.~Mikul\'a\v sek\inst{1}
        \and G.\,W.\,Henry\inst{2} \and J.~Jan\'ik\inst{1}
        \and O. Kochukhov\inst{3}\and A.~Pigulski\inst{4}\and P.~Leto\inst{5}
        \and C.~Trigilio\inst{5} \and I.~Krti\v{c}kov\'a\inst{1} \and
        T.~L\"uftinger\inst{6} \and M.~Prv\'ak\inst{1} \and A. Tich\'y\inst{1}}

%\offprints{J.~Krti\v{c}ka,\\  \email{krticka@physics.muni.cz}}

\institute{Department of Theoretical Physics and Astrophysics,
           Masaryk University, Kotl\'a\v rsk\' a 2, CZ-611\,37 Brno,
           Czech Republic
           \and
           Center of Excellence in Information Systems, Tennessee State
           University, Nashville, Tennessee, USA
           \and
           Department of Physics and Astronomy, Uppsala University, Box 516,
           751 20, Uppsala, Sweden
           \and
           Astronomical Institute, Wroc{\l}aw University, Kopernika 11, 51-622, Wroc{\l}aw, Poland
           \and
           INAF -- Osservatorio Astrofisico di Catania, Via S. Sofia 78,
           95123 Catania, Italy
           \and
           Institut f\"ur Astronomie, Universit\"at Wien,
           T\"urkenschanzstra\ss e 17, 1180 Wien, Austria}
\date{Received}

\abstract
{CU\,Vir has been the first main sequence star that showed regular radio pulses
that persist for decades, resembling the radio lighthouse of pulsars and
interpreted as auroral radio emission similar to that found in planets. The
star belongs to a rare group of magnetic chemically peculiar stars with variable
rotational period.}
{We study the ultraviolet (UV) spectrum of CU\,Vir obtained using STIS
spectrograph onboard the Hubble Space Telescope (HST) to search for the source of radio emission and to test the model of the rotational period evolution.}
{We used our own far-UV and visual photometric observations supplemented with the
archival data to improve the parameters of the quasisinusoidal long-term
variations of the rotational period. We predict the flux variations of CU\,Vir
from surface abundance maps and compare these variations with UV flux
distribution. We searched for wind, auroral, and interstellar lines in the
spectra.} 
{The UV and visual light curves display the same long-term period variations
supporting their common origin. New updated abundance maps
provide better agreement with the observed flux distribution.
The upper limit of the wind mass-loss rate is about
$10^{-12}\,M_\odot\,\text{yr}^{-1}$. We do not find any auroral lines. We
find rotationally modulated variability of interstellar lines, which is most
likely of instrumental origin.}
{Our analysis supports the flux redistribution from far-UV to near-UV and visual
domains originating in surface abundance spots as the main cause of the flux
variability in chemically peculiar stars.  Therefore, UV and optical variations
are related and the structures leading to these variations are rigidly
confined to the stellar surface. The radio emission of CU\,Vir is most
likely powered by a very weak presumably purely metallic wind, which leaves no imprint in spectra.}


\titlerunning{HST/STIS analysis of the first main sequence pulsar CU\,Vir}
\authorrunning{J.~Krti\v{c}ka et al.}
\maketitle

\section{Introduction}

The magnetic chemically peculiar star CU\,Virginis (HR\,5313, HD\,124224) is one
of the most enigmatic stars of the upper part of the main sequence. \hvezda\
shows continuum radio emission, explained as due to gyrosynchrotron process from
electrons spiraling in a co-rotating magnetosphere. According to \citet{leto06},
a stellar wind with mass-loss rate on the order of
$10^{-12}\,{M}_\odot\,\text{yr}^{-1}$ is a source of free electrons. But the
most intriguing behavior of \hvezda\ is the presence of a coherent, highly
directive radio emission at 1.4~GHz, at particular rotational phases,
interpreted as electron cyclotron maser emission \citep{trigilio,ravi,lo}. In
this interpretation, \hvezda\ is similar to a pulsar, even if the emission
process is different, resembling auroral activity in Earth and major planets
\citep{trigi11}. This type of emission is not unique among hot stars, as
evidenced by the detection of electron cyclotron maser emission from other
magnetic chemically peculiar stars \citep{to,lenc,letro}. The auroral radio
emission from stars with a dipole-like magnetic field has been modeled by
\citet{leto16}. This opens a new theoretical approach to the study of the
stellar auroral radio emission and, in particular, this method could be used to
identify the possible signature of a star-planet magnetic interaction (like the
Io-Jupiter interaction, \citealt{hezr,letomdwarf}).

However, a star with parameters that correspond to \hvezda\ is not expected to
have a stellar wind required to explain the observed continuum radio emission
\citep{metuje}. Moreover, weak winds are subject to decoupling of metals
\citep{op,kkiii,ufo}, which has not yet been observationally verified.
Consequently, the detection of ``auroral'' utraviolet (UV) lines is desirable to
further support to the current model of pulsed radio emission.

\hvezda\ belongs to the group of hot magnetic chemically peculiar stars. These
hot stars are usually slowly rotating, and possess notable outer envelopes mostly without
convective and meridional motions. This allows the radiatively driven diffusion
to be effective, and with occurrence of a global (possibly fossil) magnetic
field, it may give rise to the chemical peculiarity and consequently to
persistent surface structures \citep{mpoprad,ales}. These structures manifest
themselves in a periodic spectrum and spectral energy distribution variability,
frequently accompanied by variations of the effective magnetic field
\citep{kus,lroap,kocuvir}. The persistent surface structures (spots) differ from
transient dark solar-type spots that originate due to the suppression of
convective motion in the regions with enhanced local magnetic field and last for
short time scales only.

The detailed nature of the rotationally modulated light variability of
chemically peculiar stars remained elusive for decades. The problem became
clearer after application of advanced stellar model atmospheres calculated
assuming elemental surface distributions derived from surface Doppler mapping
based on optical spectroscopy. This enabled us to construct photometric surface
maps and to predict light variability \citep[e.g.,][]{mycuvir,prvalis}. The
result supports the current paradigm for the nature of the light variability of
chemically peculiar stars, according to which this light variability is caused
by the redistribution of the flux from the far-UV to near-UV and visible regions
due to the bound-bound (lines, mainly iron or chromium, \citealt{kodcar,molnar})
and bound-free (ionization, mainly silicon and helium, \citealt{peter,lanko})
transitions, the inhomogeneous surface distribution of elements, and modulated
by the stellar rotation. However, in \hvezda\ this model is not able to explain fully
the observed UV and optical light variability \citep{mycuvir},
particularly in the Str\"omgren $u$ color and around $1600\,$\AA\ in the UV
domain.

Moreover, being an unusually fast rotator, \hvezda\  belongs to a rare group of
magnetic chemically peculiar stars that show secular variations of the rotation
period \citep{pyperper,pyper2,mikvar,mikmon}. These variations were interpreted
as a result of torsional oscillations  \citep{krtvar}, which stem from the
interaction of an internal magnetic field and differential rotation \citep{mw}.
The variations were not, however, directly detected in the UV domain.

All these open problems connected with the atmosphere, magnetosphere, and wind 
of \hvezda\ can be adequately addressed by dedicated high-resolution UV
observations. Here we describe our HST/STIS observations aiming to get a more
detailed picture of the \hvezda\ UV light variability and its magnetosphere.

\section{Observations}
\subsection{HST and IUE ultraviolet observations}
Our HST observations were collected within the program \#14737 using the STIS
spectrograph with 31X0.05NDA slit, E140H grating in positions corresponding to
the central wavelengths 1380\,\AA\ and 1598\,\AA, and using the FUV-MAMA
detector. The observations are homogeneously distributed over \hvezda\ rotational
phase and cover the wavelength regions 1280\,--\,1475\,\AA\ and
1494\,--\,1688\,\AA\ (see Table~\ref{hst}). The observations were processed via
standard STIS pipelines and were downloaded from the MAST archive. We further
applied the median filter to reduce the noise in the spectra.

\begin{table}[t]
\caption{HST/STIS observations of \hvezda.}
\label{hst}
\centering
\begin{tabular}{cccc}
\hline
Spectrum  & Region [\AA] & HJD$_{\rm mean}$ & mean phase\\
\hline
od5y02010 & 1380 & 2\,457\,948.6933 & 0.220\\
od5y02020 & 1598 & 2\,457\,948.7062 & 0.245\\
od5y03010 & 1380 & 2\,457\,945.6472 & 0.370\\
od5y03020 & 1598 & 2\,457\,945.6602 & 0.395\\
od5y04010 & 1380 & 2\,457\,953.5934 & 0.631\\
od5y04020 & 1598 & 2\,457\,953.6067 & 0.657\\
od5y05010 & 1380 & 2\,457\,947.4351 & 0.804\\
od5y05020 & 1598 & 2\,457\,947.4480 & 0.828\\
od5y51010 & 1380 & 2\,458\,254.2392 & 0.036\\
od5y51020 & 1598 & 2\,458\,254.2522 & 0.061\\
\hline
\end{tabular}
\tablefoot{The phases $\varphi$ were calculated
according to the ephemeris Eq.\,(\ref{modelphfun}).}
\end{table}

\begin{figure}[t]
\centering \resizebox{\hsize}{!}{\includegraphics{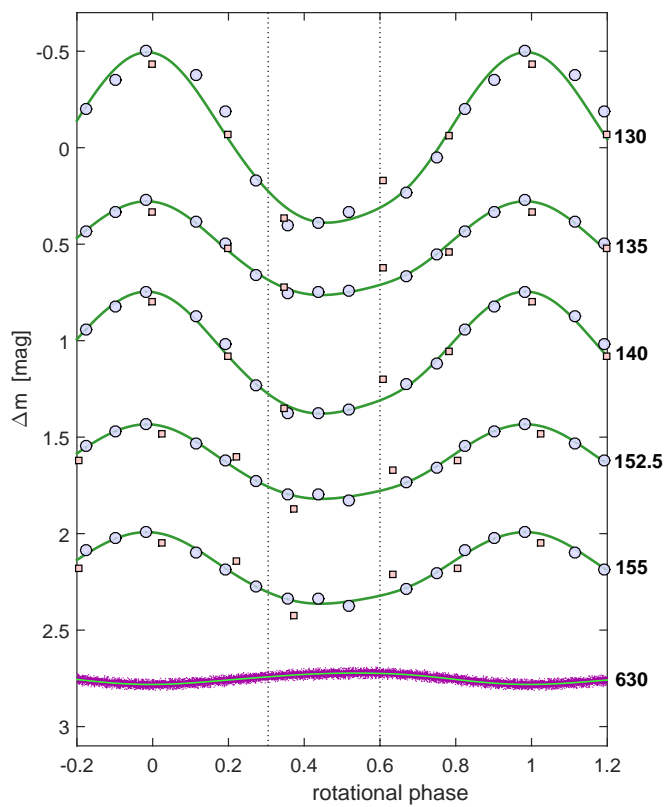}}
\caption{Comparison of light curves in far UV and red regions of \hvezda\
spectrum. Blue circles and red squares denote synthetic magnitudes derived from
IUE and STIS spectra and lilac points are the measurements from SMEI satellite.
Wavelengths are expressed in nanometers. Phases are calculated using ephemeris
Eq.~(\ref{modelphfun}) applying parameters given in Table\,\ref{ephemeris}.
Green lines are the fits of the light curve phenomenological model
Eq.~(\ref{fena}), assuming two photometric spots with photocenters at the phases
0.303 and 0.598 (denoted using vertical black dotted lines). Light curves were
plotted relative to the mean value of the fits and vertically shifted to
separate individual plots.}
\label{LCall}
\end{figure}

Observations from the HST nicely cover the rotation cycle of the star (see
Fig.~\ref{LCall}). The observations close to the phase $\varphi\approx0$
correspond to visual minimum and to UV maximum. This is also the phase of the
start of XMM-Newton X-ray observation ($\varphi\approx0.07$,
\citealt{takyjsemnasel}). The spectra od5y02010 and od5y02020 correspond to the
period of X-ray quiescence roughly 7\,ks after the start of observation. The
spectrum od5y03010 was obtained close to the $a$ peak of radio emission, which
is observed around the phases $0.29-0.35$ \citep{trigi08}. This is also approximately
the period of maximum strength of Si lines, which appears roughly around the
phase $\varphi\approx0.35$ \citep{kus} and which corresponds to strong X-ray
flux roughly 15\,ks after the start of XMM-Newton observations. The maximum
strength of Fe lines and optical maximum is observed slightly later around the
phase $0.5$. The spectrum od5y04020 marks the end of XMM-Newton observations.
The peak $b$ of radio emission is located around $\varphi=0.71-0.77$, close to
which the spectrum od5y05010 was obtained. This also roughly marks the start of
Chandra observations ($\varphi=0.76$). These observations end at
$\varphi=0.40$.

We supplemented the HST \hvezda\ observations with IUE data \citep[PI:
M.~Maitzen, see][hereafter \citetalias{mycuvir}]{sokolpan,mycuvir}. To obtain
the UV photometry, we extracted IUE observations of \hvezda\ from the INES
database using the SPLAT package \citep[][see also \citealt{pitr}]{splat}. We
downloaded low-dispersion data originating from high-dispersion large aperture
spectra in the domains 1250--1900~\AA\ (SWP camera) and 2000--3000~\AA\ (LWR
camera). For the spectroscopic analysis, we used high-dispersion IUE data
downloaded from the MAST archive. We compared the absolute flux normalization of
HST and IUE spectra concluding that these flux normalizations are consistent
(within small differences which can be attributed to calibration).

We used the HST and IUE SWP fluxes $F(\lambda)$ to calculate UV
magnitudes (see Fig.~\ref{LCall})
\begin{equation}
\label{magrov}
m_c=-2.5\log\left[\int_0^{\infty}\mathit\Phi_c(\lambda) F(\lambda)\,\de\lambda
\right],
\end{equation}
where
$\mathit\Phi_c(\lambda)=\zav{\sqrt\pi\sigma}^{-1}\exp\zav{-{(\lambda-c)^2}/{\sigma^2}}$ is a normalized Gauss function centered on the wavelength $c$. The central wavelengths of individual Gauss filters cover HST spectra, $c=1300\,\AA$, $1350\,\AA$,
$1400\,\AA$, $1525\,\AA$, and $1550\,\AA$. The dispersion was selected
$\sigma=10\,\AA$ for $c=1300\,\AA$ and $1525\,\AA$, and $\sigma=25\,\AA$ for the
remaining ones.
Figure~\ref{LCall} shows that the IUE and HST light curves are very
similar with small differences most likely due to instrumental calibration.

We derived the synthetic photometry from IUE LWR spectra also using
Eq.~\eqref{magrov} \citep[see][]{mikvar}. The UV synthetic photometry of HST and
IUE was further supplemented by the data derived from OAO II satellite
\citep{molnarwu}.

\subsection{Standard photometry}

In addition to the 326 photometric measurements derived from the UV spectra
taken by HST and IUE and adopted from \citet{molnarwu}, we also used 35\,803
other available photometric data obtained in the time interval 1972 -- 2018 by
various observers in the region 330--753 nm using standard photometric
techniques \citep[see][for the list]{mikmon}. We supplemented these data with two
additional sets of observations.

From 2010 February through 2018 June, we acquired 1191 photometric
observations in the Johnson $V $ band and 1327 in the Johnson  $B$ band, with Tennessee
State University's T3 0.40~m automatic photoelectric telescope (APT) at
Fairborn Observatory in the Patagonia Mountains of southern Arizona
\citep{h99}.  Each observation in each filter represents the mean of three
differential measurements of CU Vir bracketed by four measurements of its
comparison star HD~125181.  Details on the acquisition and reduction of the
observations can be found in our paper on HR~7224 \citep{myhr7224}.

%\subsubsection{SMEI photometry}

The space photometry of {\hvezda} is available  from the Solar Mass Ejection
Imager (SMEI) experiment \citep{2003SoPh..217..319E,2004SoPh..225..177J}. The
experiment was placed on-board the Coriolis spacecraft and aimed to
measure sunlight scattered by free electrons of the solar wind. A by-product
of the mission is the photometry of bright stars all over the sky, available
through the University of California San Diego (UCSD) web
page\footnote{http://smei.ucsd.edu/new\_smei/index.html}. The SMEI passband
covers the range between 450 and 950~nm \citep{2003SoPh..217..319E}. The
photometry of stars was obtained using a technique developed by
\cite{2007SPIE.6689E..0CH}. The SMEI stellar time series are affected by
long-term calibration effects, especially a repeatable variability with a
period of one year. They also retain unwanted signals from high energy particle
hits and aurorae. Nevertheless, the SMEI observations are well suited for
studying variability of bright stars primarily because they cover an extended  
interval --  2003 to 2010.

The raw SMEI UCSD photometry of {\hvezda} was first corrected for the one-year
variability by subtracting an interpolated mean light curve. This curve was
obtained by folding the raw data with the period of one year, calculating
median values in 200 intervals in phase, and then interpolating between them.
In addition, the worst parts of the light curve and outliers were removed.
Then, a model consisting of the sinusoid with rotational frequency of {\hvezda}
and its detectable harmonics was fitted to the data. The low-frequency
instrumental variability was filtered out by subtracting trends that were calculated
using residuals of the fit and removing outliers using $\sigma$-clipping. The
trends were approximated by calculating averages in time intervals which were
subsequently interpolated. The final SMEI light curve consists of 19\,226
individual data points.

\subsection{Optical spectroscopy}

Additional optical spectra were obtained at Astronomical observatory in Piwnice
(Poland). We used a mid-resolution ech\`{e}lle spectrograph ($R\sim15\,000$)
connected via optical fiber to the 90\,cm diameter Schmidt-Cassegrain telescope.
Spectra were observed during two weekly runs in 2010 (March 3\,--\,10,
April 4\,--\,8) and during one night in 2011 (February 10). Th-Ar calibration
spectrum was taken before and after each scientific image. For reduction we used
standard IRAF\footnote{IRAF is distributed by NOAO, which is operated by AURA,
Inc., under cooperative agreement with the National Science Foundation.}
routines (bias, flat, and wavelength calibration). In total we collected 59
spectra, which cover wavelength regions from 4350\,--7160~\AA~and all rotational
phases of the star.

\section{Long-term rotational period variations}

The method used to monitor \hvezda\ long-term rotational period variations is
based on the phenomenological modeling of phase function $\vartheta(t)$ (the sum of the phase and epoch) and periodic or quasiperiodic phase curves of tracers of the rotation (typically light curves), as the functions of $\vartheta(t)$. The phase function is connected with the generally variable rotational period $P(t)$ through the basic relations \citep[see][]{mikori,mikmon}:
\begin{equation}\label{phasefun}
\displaystyle \dot{\vartheta}=\frac{1}{P(t)};\quad \vartheta(t)=\int_{M_0}^t \frac{\mathrm{d}\tau}{P(\tau)};\quad \theta(\vartheta)=\int_0^\vartheta P(\zeta)\,\mathrm{d}\zeta.
\end{equation}
The phase function $\vartheta(t)$ is a smooth, monotonically rising function of
time. The relation for its inversion function $\theta(\vartheta)$ can be derived
from the same basic differential equation. Using the inversion phase function,
we can predict, for example, times of maxima or other important moments of the phase variations. 

\subsection{Model of light curves}

The light curves, obtained at various times in the filters of different
effective wavelengths $\lambda_{\mathrm{eff}}$ from 130 to 753~nm, provide a
massive amount of information about \hvezda\ rotation and its long-term
variations. There is a general expectation \citepalias{mycuvir} that the shapes
of \hvezda\ light curves strongly depend on the effective wavelengths (see
Fig.\,\ref{LCall}). Nevertheless, the careful inspection revealed that all of
the observed light curves can be satisfactorily expressed as a linear
combination of only two symmetric basic profiles with centers at phases
$\vartheta_{0j}$ and half-widths $d_j$:
\begin{eqnarray}\label{fena}
&\displaystyle
F(\lambda_{\text{eff}},\vartheta)=\sum_{j=1}^{2}A_j(\lambda_{\text{eff}}) \,\szav{\exp\hzav{1\!-\!\cosh\zav{\frac{\Delta \varphi_j}{d_j}}}-2.29\,d_j};\nonumber\\
&\mathrm{where}\quad\Delta\varphi_j=(\vartheta-\varphi_{0j})- \mathrm{round}(\vartheta-\varphi_{0j}),
\end{eqnarray}
where $F(\lambda_{\text{eff}},\vartheta)$ is the model prediction of the
difference of the magnitude with respect to its mean value as a function of the
wavelength $\lambda_{\mathrm{eff}}$ and phase function $\vartheta$, $A_1$ and
$A_2$ are amplitudes depending on wavelengths, $d_1$ and $d_2$ are the
half-widths of the basic profiles expressing their sizes, and $\varphi_{01}$ and
$\varphi_{02}$ are the phases of their centers. This phenomenological model
quantifies the fact known from our previous studies that the areas with abnormal spectral energy distribution of the flux on the surface are
concentrated around just two prominent asymmetrically distributed conglomerations
\citepalias{mycuvir}.

The phase of the center of the first spot was fixed at $\varphi_{01}=0.598$,
while the location of the center of secondary photometric spot
($\varphi_{02}=0.3029(6)$), as well as the basic profile half-widths,
$d_1=0.1774(8)$ and $d_2=0.1491(6)$, were determined by iterativelly fitting
observations. The zero phase corresponds to the light curve minimum in
near UV and optical regions and to the maximum in far UV (see
Fig.\,\ref{LCall}).

\subsection{Model of phase function and period variations}\label{modelphfunkap}

\begin{figure}[t]
\centering
\resizebox{\hsize}{!}{\includegraphics{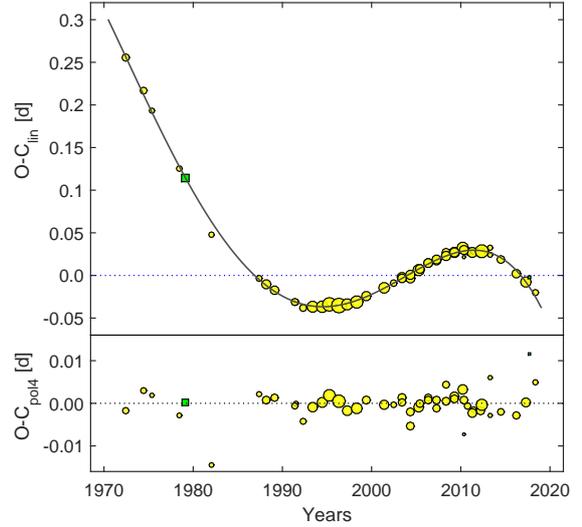}}
\caption{\emph{Upper plot}: $\mathit{O\!-\!C}_{\rm{lin}}$ versus time in years.
 $O$ is the moment of the zero phase and $C_{\rm{lin}}=M_{01}+P_{1}\times
E$.  $M_{01}$ and $P_{1}$ are the parameters of the linear approximation of
$O(E)$ behavior, $E$ is the number of rotational cycles elapsed from the
initial epoch $M_{01}$ and $P_1$ is the mean rotational period in the interval
1970--2018; $M_{01}=2\,452\,650.9048$ and $P_1=0.52070819$\,d. Nonlinearity of
$\mathit{O\!-\!C}_{\rm{lin}}$ is a result of variable rotational period $P$.
$\mathit{O\!-\!C}_{\rm{lin}}$ is fitted by a fourth-order polynomial (solid
line). \emph{Bottom plot}: The residuals $\mathit{O\!-\!C}_{\rm{pol4}}$ show
that the model ensures the accuracy of ephemeris $ 0.0025\,\text{d}$ or $0.005$
in phase. Optical photometry is denoted using yellow circles and UV
spectrophotometry derived from observed fluxes using green squares. Sizes of the
symbols are inversely proportional to the square root of the uncertainties of
the determinations of $O$ times derived from individual data sources.}
\label{OC}
\end{figure}

\begin{table}[t]
\caption{Parameters of \hvezda\ ephemeris (Eq.~\eqref{modelphfun}).}
\label{ephemeris}
\centering
\begin{tabular}{ccc}
\hline
$M_0$ & (HJD)& 2\,452\,650.8991(6)\\
$P_0$ &[d]& 0.520\,716\,59(8)\\
$\dot{P}_0$& &4.9(2)$\,\times\,10^{-10}$\\
$\ddot{P}_0$& $[\mathrm{d}^{-1}]$ &$-1.762(11)\,\times\,10^{-12}$\\
$\dddot{P}_0$ &$[\mathrm{d}^{-2}]$& $-3.23(5)\,\times\, 10^{-16}$\\
\hline
\end{tabular}
\end{table}

The primary purpose of the proposed definition of the phase function model is
finding as simple as possible ephemeris guaranteeing the accuracy in the phase
of at least 0.01 in 1970 -- 2018. In this time interval, there were obtained
36\,129 photometric observations or photometric measurements derived from UV
spectra taken by IUE and HST. The data create 177 individual light curves in 37
filters, covering substantial part of \hvezda\ electromagnetic spectrum. This volume of data is quite sufficient for  approximation of the dimensionless phase curve $\vartheta(t)$ and its inversion function $\theta(\vartheta)$ (with a time dimension) by the fourth order polynomial in the form of the Taylor decomposition:
\begin{align}
&\vartheta_0=\textstyle{\frac{t-M_0}{P_0}};\quad  \vartheta=\vartheta_0-\textstyle{\frac{1}{2!}}\dot{P}_0\vartheta_0^2-
\textstyle{\frac{1}{3!}}P_0\ddot{P}_0\vartheta_0^3-
\textstyle{\frac{1}{4!}}P_0^2\dddot{P}_0\vartheta_0^4; \nonumber\\
&\theta(\vartheta)\!=\!M_0\!+\!P_0\vartheta\!+
\!\textstyle{\frac{1}{2!}}P_0\dot{P}_0\vartheta^2+
\textstyle{\frac{1}{3!}}P_0^2\ddot{P}_0\vartheta^3\!+\!
\textstyle{\frac{1}{4!}}P_0^3\dddot{P}_0\vartheta^4; \label{modelphfun}\\
&\varphi=\mathrm{frac}({\vartheta});\quad E=\mathrm{floor}({\vartheta});\label{epoch}\\
&P(t)\doteq P_0+P_0\,\dot{P}_0\,\vartheta_0+\textstyle{\frac{1}{2!}}P_0^2\,\ddot{P}_0\,\vartheta_0^2+
\textstyle{\frac{1}{3!}}P_0^3\dddot{P}_0\vartheta_0^3, \label{instP}
\end{align}
where $\vartheta_0$ is an auxiliary dimensionless quantity, $M_0$ is the origin
of phase function, which was placed close to the weighted center of the
observations, $P_0,\,\dot{P}_0,\,\ddot{P}_0,$  and $\dddot{P}_0$ are the
instantaneous period and its time derivatives up to third degree at the time
$t=M_0$, $\varphi$ is the usual phase, and $E$ is the epoch.
We selected the Taylor expansion \eqref{modelphfun} as, contrary
to the harmonic expansion \citep{krtvar}, it allowed us to directly calculate
the phase and its inversion. Using relations in Eqs.~(\ref{modelphfun}) and
(\ref{epoch}) we can calculate the phase function $\vartheta$, the epoch $E$,
and the phase $\varphi$, for any HJD moment $t$ within the
interval 1972 -- 2020. Alternatively, knowing the particular phase function
$\vartheta$ we can estimate corresponding time $\theta(\vartheta)$. Changes
of the instantaneous period $P(t)=P(\vartheta_0)$ can be evaluated
for any time $t$ using relation~(\ref{instP}).

\subsection{Model solution and brief discussion of results}

All 259 free parameters of our phenomenological model of \hvezda\ light curves
and their corresponding uncertainties were determined together by robust regression (RR) as implemented in \citet{mikRR}, which exploits the well-established procedures of the standard weighted least squares method and eliminates the influence of outliers. Instead of the usual $\chi^2$, we minimized the modified quantity $\chi^2_{\rm r}$, defined as follows:
\begin{eqnarray}
&\Delta y_i=y_i-F(t_i,\boldsymbol{\gamma}),\\
& \chi^2_{\rm r}=\sum_{i=1}^n\,\zav{\frac{\Delta y_i} {\sigma_{\mathrm{r}i}}}^2; \quad \mathrm{where}\ \ \sigma_{\mathrm{r}i}=\sigma_i\ \exp\hzav{\frac{1}{2}\zav{\frac{\Delta y_i}{4\,\sigma_i}}^4},\\
&\chi^2_{\mu}=1.06\ \frac{\chi^2_{\mathrm{r}}}{n_{\rm r}-g};\quad \mathrm{where}\ \
n_{\rm r}=1.02\ \frac{\sum\,\sigma_{\mathrm{r}i}^{-2}}{\sum\,{\sigma_{i}}^{-2}},
\label{chimi}\\
& \delta\gamma_{k}=\sqrt{\chi^2_{\mu}\,\bold{H}_{kk}},\label{delta}
\end{eqnarray}
where $\Delta y_i$ is the difference between the observed $i$-th measurement
$y_i$ and the model prediction $F(t_i,\boldsymbol{\gamma})$, which is the
function of the time of the measurement $t_i$ and the vector of the free model
parameter $\boldsymbol{\gamma}$, $\sigma_i$ is the estimate of the uncertainty
of determination of the $i$-th measurement, while $\sigma_{\mathrm{r}i}$ is a RR
modified value of $\sigma_i$. The estimate of the common relative $\chi^2_{\mu}$
and the number of measurements without outliers $n_{\rm r}$ are given by Eq.~(\ref{chimi}).

The formal uncertainties of the found parameters $\delta\boldsymbol{\gamma}$
were determined by the relation Eq.~(\ref{delta}), where $\bold{H}$ is a
covariant matrix of the system of equations. The most interesting parameters and
their functions are listed in Table\,\ref{ephemeris}.

The comparison between observation and model predictions in Fig.\,\ref{OC}
undoubtedly documents the high fidelity of the chosen model, which yields the
phase function ephemeris describing observations with the accuracy better than
0.005 in the interval 1972--2018. The derived period variations can be
adequately fitted also by the harmonic polynomials instead of the Taylor
expansion \eqref{modelphfun} and interpreted as due to the torsional
oscillations \citep{krtvar}.

\section{Variations of the UV spectrum}

\subsection{Simulation of the spectral variability}

\begin{table}[t]
\caption{\hvezda\ parameters from spectroscopy.}
\label{hvezda}
\centering
\begin{tabular}{lc}
\hline
\multicolumn{2}{c}{\citealt{kus}}\\
\hline
Effective temperature ${{T}_\mathrm{eff}}$ & ${13\,000}$\,K \\
Surface gravity ${\log g}$ (cgs) & ${4.0}$ \\
Inclination ${i}$ & ${30^\circ}$ \\
Helium abundance&$-3.2<\varepsilon_\text{He}<-1$ \\
Silicon abundance& $-4.6<\varepsilon_\text{Si}<-2.3$ \\
Chromium abundance& $-6.7<\varepsilon_\text{Cr}<-4.4$ \\
Iron abundance&$-5.5<\varepsilon_\text{Fe}<-3.5$  \\
\hline
\multicolumn{2}{c}{\citealt{kocuvir}}\\
\hline
Effective temperature ${{T}_\mathrm{eff}}$ & ${12\,750}$\,K \\
Surface gravity ${\log g}$ (cgs) & ${4.3}$ \\
Radius $R_*$ & $2.06\,R_\odot$\\
Mass $M$ & $3.06\,M_\odot$\\
Inclination ${i}$ & ${46.5^\circ}$ \\
Silicon abundance& $-5.7<\varepsilon_\text{Si}<-2.3$ \\
Iron abundance&$-4.7<\varepsilon_\text{Fe}<-3.1$  \\
\hline
\end{tabular}
\end{table}

The calculation of the model atmospheres and spectra is the same as described in
\citetalias{mycuvir}. We used the TLUSTY code for the model atmosphere
calculations \citep{tlusty,hublaj,hublad,lahub} with the atomic data taken from
\citet{bstar2006}. In particular, the atomic data for silicon and iron are based
on \citet{mendo}, \citet{maslo93}, \citet{kur22}, \citet{nah96}, \citet{nah97},
\citet{bau97}, and \citet{bau96}, and for other elements on \citet{top1},
\citet{topf}, \citet{toptul}, \citet{topp}, \citet{toph}, and \citet{napra}. We
prepared our own ionic models for chromium (\ion{Cr}{ii}--\ion{Cr}{v}) using
data taken from Kurucz (2009)\footnote{http://kurucz.harvard.edu}.

We used two different sets of surface abundance maps. \citet{kus} derived
abundance maps of helium, magnesium, silicon, chromium, and iron.
\citet{kocuvir} derive abundance maps of silicon and iron only, but using an
updated version of the inversion code that accounts also for the magnetic field.
For each set of maps (from \citet{kus} and \citet{kocuvir}) we adopted corresponding
effective temperatures and surface gravities derived from spectroscopy (see
Table~\ref{hvezda}) and assumed the range of abundances that covers values in
individual maps as given in Table~\ref{hvezda}. Here the abundances are given
relative to hydrogen, that is,
$\varepsilon_\text{el}=\log\zav{N_\text{el}/N_\text{H}}$. We neglected the
influence of inhomogeneous surface distribution of magnesium due to its low
abundance. For each set of maps we used inclination that was derived from
corresponding mapping (see Table ~\ref{hvezda}). We used the solar abundance of
other elements \citep{asgres}.

For the calculation of synthetic spectra we used the SYNSPEC code. The spectra
were calculated for the same parameters (effective temperature, surface gravity,
and chemical composition) as the model atmospheres. We also took into account
the same transitions (bound-bound and bound-free) as for the model atmosphere
calculations. This is important for iron line transitions, for which many lines
without experimental values significantly influence the spectral energy
distribution. The remaining line data are taken from the line lists available at
the TLUSTY web page. We computed emergent specific intensities for $20$
equidistantly spaced values of $\mu=\cos\theta$, where $\theta$ is the angle
between the normal to the surface and the line of sight.

The model atmospheres and the emergent specific intensities
$I(\lambda,\theta,\varepsilon_\text{He},\varepsilon_\text{Si},\varepsilon_\text{Cr},\varepsilon_\text{Fe})$
were calculated for a four-parametric grid of helium, silicon, chromium, and
iron abundances (see Table~\ref{esit}). The lowest abundances of individual
elements are omitted from the grid. Our test showed that this restriction does
not influence the predicted light curves.

\begin{table}[t]
\caption{Individual abundances $\varepsilon_\text{He}$, $\varepsilon_\text{Si}$,
$\varepsilon_\text{Cr}$, and $\varepsilon_\text{Fe}$ of the model grid.}
\label{esit}
\centering
\begin{tabular}{lrrrrrrr}
\hline
He& $-2.0$ &$-1.0$\\%& $-0.5$& $0.0$& $0.5$& $1.0$& $1.5$ \\
Si& $-4.75$& $-4.25$& $-3.75$& $-3.25$& $-2.75$ & $-2.25$\\
Cr& $-6.4$& $-5.9$& $-5.4$& $-4.9$& $-4.4$\\
Fe & $-5.4$& $-4.9$& $-4.4$& $-3.9$& $-3.4$ &$-3.1$ \\
\hline
\end{tabular}
\end{table}

The radiative flux at the distance $D$ from the star with radius $R_*$ is
\citep{hubenymihalas}
\begin{equation}
\label{vyptok}
F_\lambda=\zav{\frac{R_*}{D}}^2\intvidpo I(\lambda,\theta,\Omega)
\cos\theta\,\text{d}\Omega.
\end{equation}
The intensity $I(\lambda,\theta,\Omega)$ at each surface point with
spherical coordinates $\Omega$ is obtained by means of
interpolation between the emergent specific intensities
$I(\lambda,\theta,\varepsilon_\text{He},\varepsilon_\text{Si},\varepsilon_\text{Cr},\varepsilon_\text{Fe})$
calculated from the grid of synthetic spectra (Table~\ref{esit}) taking into
account the Doppler shifts due to the stellar rotation.

\subsection{Comparison with the observed variations}

\begin{figure*}[tp]
\centering \resizebox{\hsize}{!}{\includegraphics{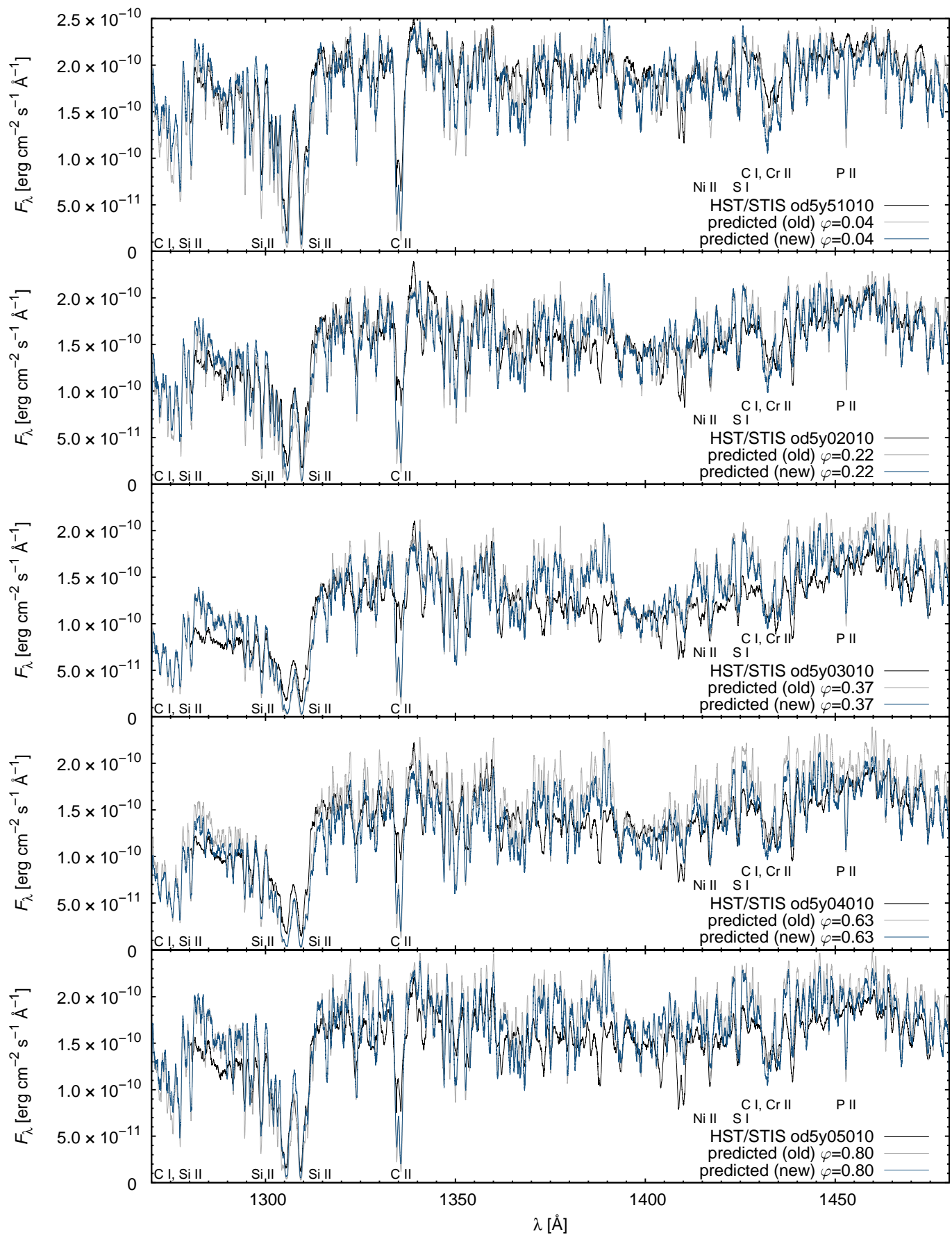}}
\caption{Predicted and observed (HST, black line) flux in selected phases for
the wavelength range 1270 -- 1480\,\AA. The gray line denotes spectra calculated
using \citet{kus} surface abundance distribution and blue line denotes spectra
calculated using \citet{kocuvir} surface abundance distribution. Individual
strong lines and iron line blends are identified.}
\label{hst1380}
\end{figure*}

\begin{figure*}[tp]
\centering \resizebox{\hsize}{!}{\includegraphics{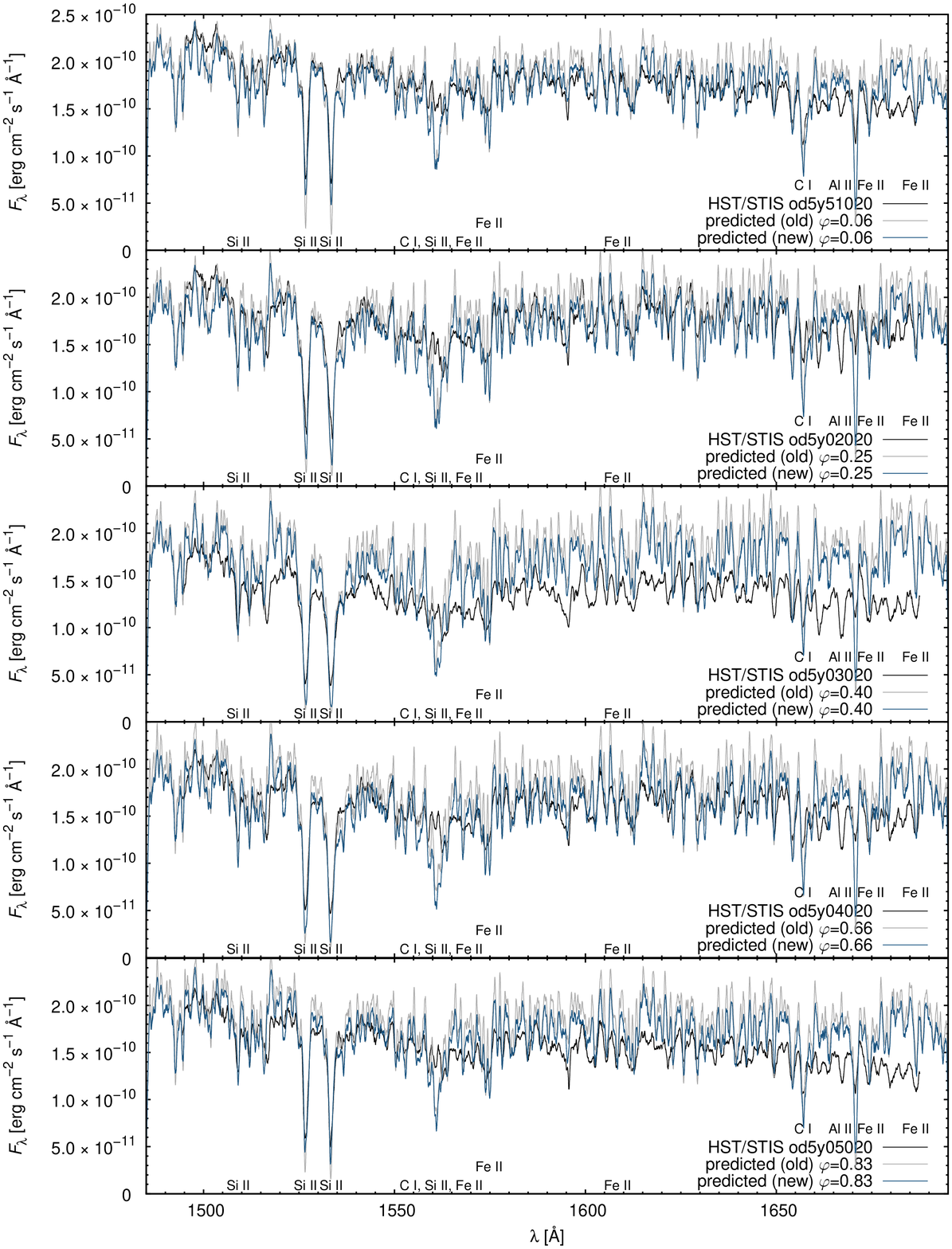}}
\caption{Same as Fig.~\ref{hst1380}, except for the wavelength range 1485 --
1695\,\AA.}
\label{hst1598}
\end{figure*}

We compared flux variations predicted using Eq.~\eqref{vyptok} with flux
variations observed by HST/STIS for individual rotational phases $\varphi$. The
predicted fluxes were scaled by the $(R_*/D)^2$ ratio derived in
\citetalias{mycuvir} from the IUE spectra.

The fluxes calculated from the two sets of abundance maps show a good agreement
with observed spectra for the spectral region around 1280\,--\,1475\,\AA\ (see
Fig.~\ref{hst1380}) especially for the rotational phase $\varphi\approx0$, when
the regions with low abundance of iron and silicon are visible. This spectral
region is dominated by silicon bound-free and iron line transitions.

%For phases
%$\varphi\approx0.2$, $\varphi\approx0.6$, and $\varphi\approx0.8$, when the
%lines of silicon and iron are strongest, the flux predicted from the \citet{kus}
%abundance maps is slightly higher than the predicted flux. However, due to
%larger spot with silicon overabundance, the surface abundance distribution of
%\citet{kocuvir} nicely fits the observed spectra even during the phases
%$\varphi\approx0.2$ and $\varphi\approx0.8$. This shows that a new generation of
%inversion codes that include also the magnetic field provide results which are
%quantitatively more reliable.

The two sets of abundance maps provide similar results for the 1494\,--\,1688\,\AA\
wavelength region, where the influence of silicon is weaker (see
Fig.~\ref{hst1598}). The predicted fluxes nicely agree with observations for
phases $\varphi\approx0$ and $\varphi\approx0.2$, while the predicted fluxes are
slightly larger than the observations for $\varphi\approx0.4$,
$\varphi\approx0.6$, and $\varphi\approx0.8$. A detailed inspection of fluxes
shows that the new maps agree with observations even slightly better during the
phases $\varphi\approx0$, $\varphi\approx0.2$, and $\varphi\approx0.6$ in the
interval $1600-1650\,$\AA. The remaining disagreement is likely caused by
missing opacity due to some additional element(s). We tested the flux
distribution calculated with enhanced abundance of light elements with $Z<30$
and we have found that a missing opacity is possibly due to chromium. The
predicted flux distribution for $\epsilon_\text{Cr}=-4$ nicely matches the
observations for $\varphi\approx0.4$. Chromium was mapped by \citet{kus}, but
new maps based on more detailed spectroscopy could possibly give higher chromium
abundance in the spots.

We also tested new line data from the VALD database \citep{vald1,vald2,vald3}.
This led to an improvement of the agreement in some specific lines, but caused
disagreement in others. Consequently, part of the discrepancy between the
predicted and observed spectra can be probably attributed to the imprecise line
data.

In addition to the line variability of elements studied in optical (silicon,
chromium, and iron), the UV spectra also show carbon line variability. Carbon is
roughly depleted by a factor of ten with respect to the solar value,
consequently it is not expected to significantly affect the flux distribution.
We also tested the presence of lines of heavy elements, with atomic number
$Z\leq92$, in the spectra of \hvezda. We used our original line list and the
line list downloaded from the VALD database \citep{vald1,vald2,vald3}. We find no evidence for the presence of lines of elements having $Z>30$ in
the spectra of \hvezda.

\section{Stellar wind of \hvezda}

The explanation of \hvezda\ continuum radio emission by gyrosynchrotron process
requires a wind mass-loss rate on the order of
$10^{-12}\,{M}_\odot\,\text{yr}^{-1}$ \citep{leto06}. However, wind models show
that the radiative force is too weak to accelerate the wind for \hvezda's
effective temperature \citep{metuje}. To understand this inconsistency, we
provide additional wind tests and search for the wind signatures in the \hvezda\
spectra.

We used our METUJE wind code \citep{metuje} to test if the radiative force
overcomes gravity in the envelope of \hvezda\ and to calculate the wind spectra.
Our wind code solves the hydrodynamic equations (equations of continuity,
motion, and energy) in a stationary spherically symmetric stellar wind. The
radiative transfer equation is solved in the comoving frame (CMF). The code
allows us to treat the atmosphere and wind in a unified (global) manner
\citep{cmfkont}, but for the present purpose we modeled just the wind with the
METUJE code, and we derived the base flux from the TLUSTY model atmosphere model
with $\epsilon_\text{He}=-1$, $\epsilon_\text{Si}=-3.25$,
$\epsilon_\text{Cr}=-5.4$, and $\epsilon_\text{Fe}=-3.1$. To enable better
comparison with the observed spectra, the TLUSTY flux in the regions of strong
wind lines was replaced by the flux predicted from inhomogeneous surface
abundances for phase $\varphi=0.30$.

For the wind tests, we assumed fixed wind density and velocity and compared the
radiative force with gravity \citep{metuje}. We tested mass-loss rates in the
range of $10^{-14}-10^{-10}\,{M}_\odot\,\text{yr}^{-1}$ for mass-fractions of
the heavy elements $Z$ relative to the solar value $Z_\odot$ in the range
$Z=Z_\odot - 10^3Z_\odot$ scaling abundances of all elements heavier than helium
by the same value. The radiative force is typically one or two orders of
magnitude lower than the gravitational acceleration (in absolute values). The
radiative
force overcomes gravity only for the most metal rich model with $Z=10^3Z_\odot$,
which is higher than the fraction of silicon and iron found in metal rich
regions on the \hvezda\ surface (see Table~\ref{hvezda}). Consequently, we find
no wind in \hvezda\ from our theoretical models. This is consistent with results
of \citet{babelb}, who also found no wind for stellar parameters corresponding
to \hvezda. A weak, purely metallic wind with $\dot
M\approx10^{-16}\,{M}_\odot\,\text{yr}^{-1}$ is still possible for such
low-luminosity stars \citep{babela}.

\begin{figure}[tp]
\centering \resizebox{\hsize}{!}{\includegraphics{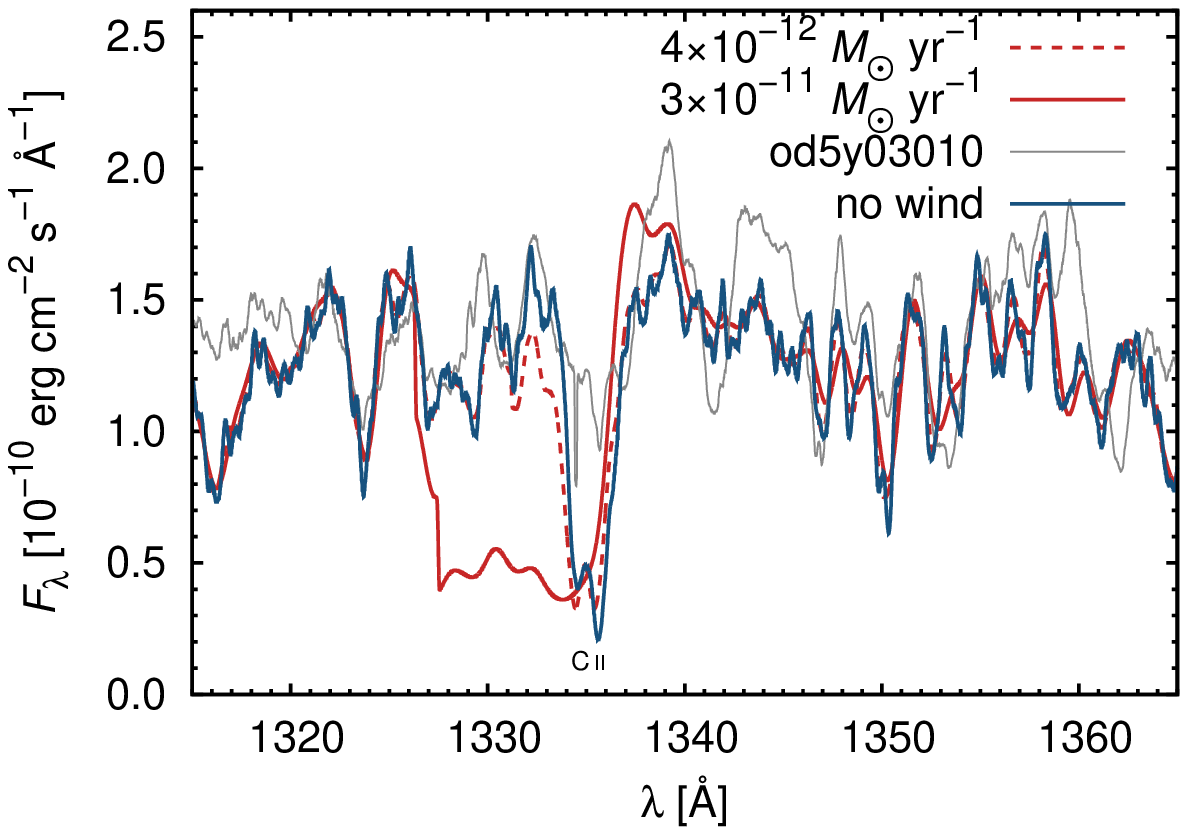}}
\centering \resizebox{\hsize}{!}{\includegraphics{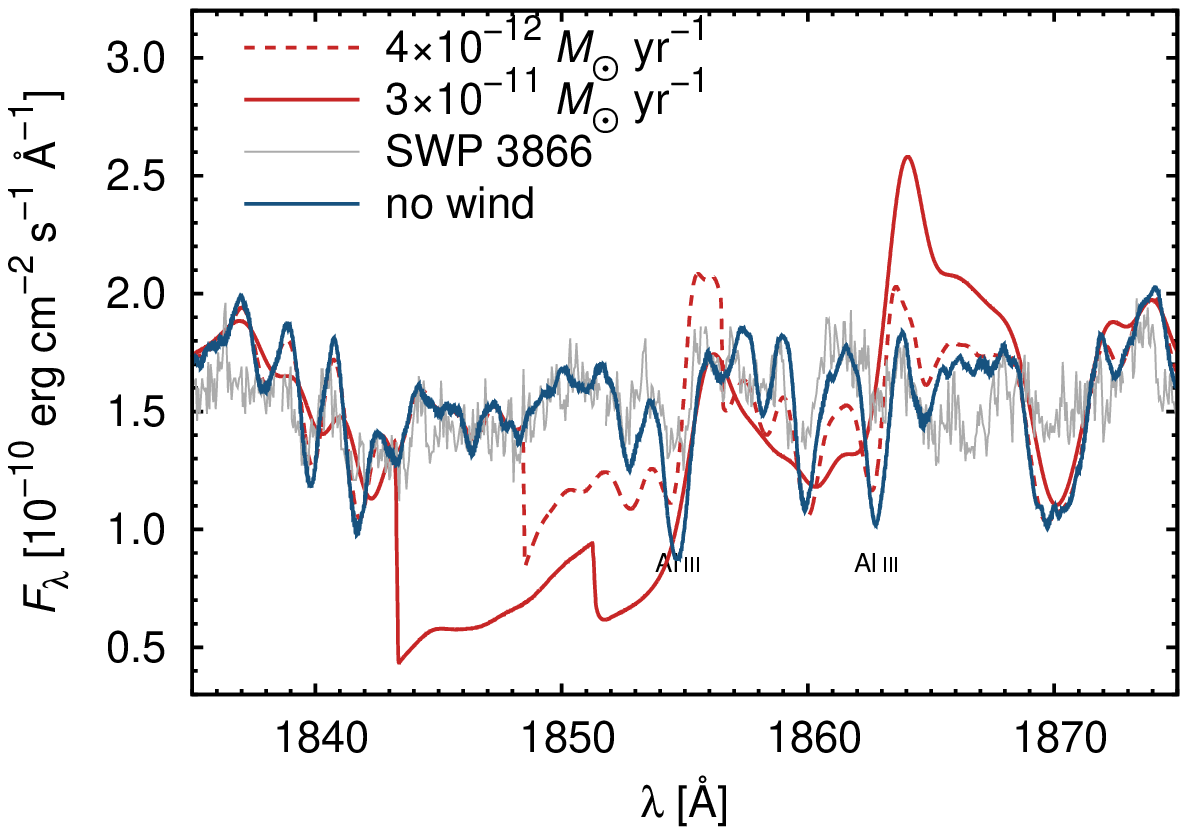}}
\centering \resizebox{\hsize}{!}{\includegraphics{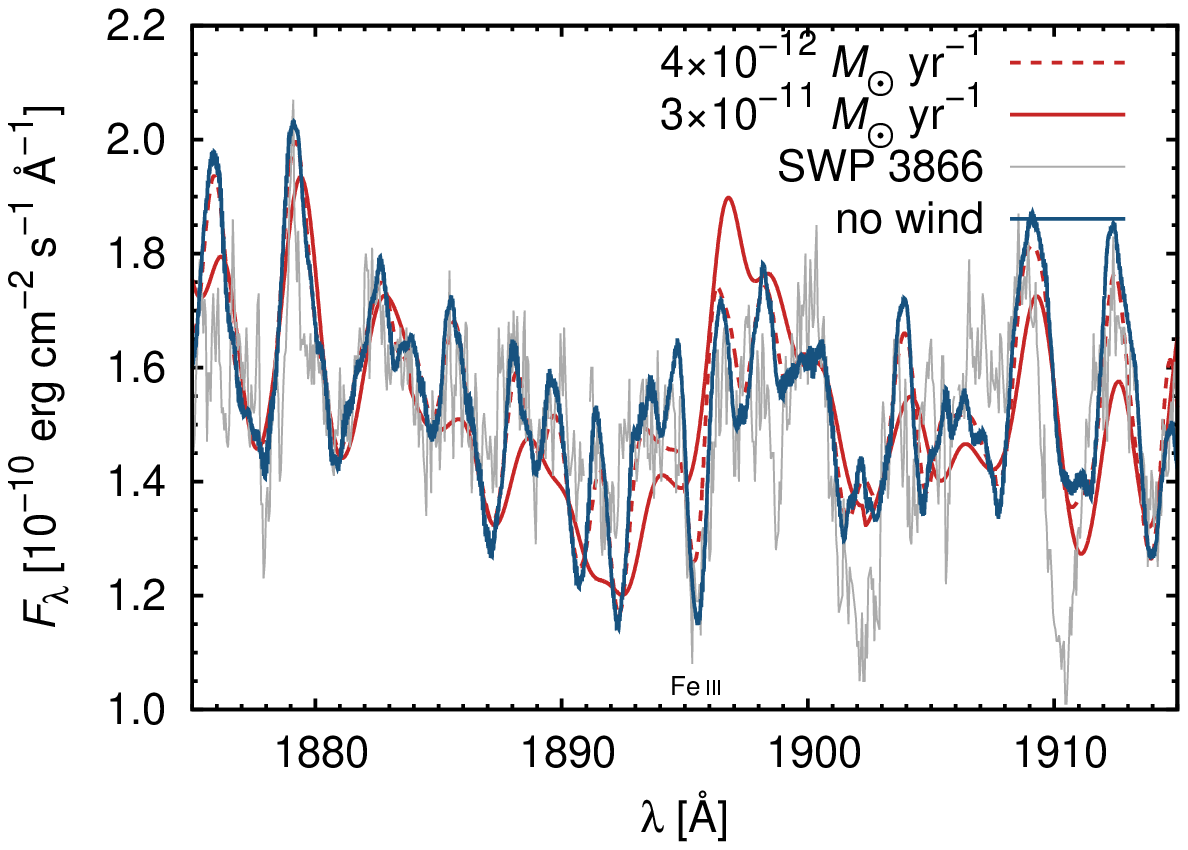}}
\caption{Predicted wind spectra for two values of mass-loss rates (red lines)
in comparison with spectra without wind (blue line) and observed spectra
(gray line).}
\label{cuvir_met}
\end{figure}

Spectral lines originating in the wind have in general complex line
profiles, which depend on the wind mass-loss rate, velocity law, and ionization
balance. Especially at low wind densities, the wind line profiles could be
hidden in blends of numerous UV lines. Therefore, to determine which wind lines
could possibly be found in the observations, we used our METUJE code to
predict the wind spectra. We assumed $Z=10^2Z_\odot$, which roughly corresponds
to the most silicon and iron rich surface layers and artificially multiplied
the radiative force by a factor of 3--10 to get wind model to converge with
different wind mass-loss rates. Below we discuss  further two representative
wind models with a mass-loss rate $\dot
M=4\times10^{-12}\,{M}_\odot\,\text{yr}^{-1}$ and terminal velocity
$v_\infty=1010\,\kms$ and a model with $\dot
M=3\times10^{-11}\,{M}_\odot\,\text{yr}^{-1}$ and $v_\infty=1850\,\kms$.

We find that the strongest wind features appear in the \ion{C}{ii} lines at
1335\,\AA\ and 1336\,\AA\ and in the \ion{Al}{iii} lines at 1855\,\AA\ and
1863\,\AA\ (Fig.~\ref{cuvir_met}). Aluminum lines are predicted to show strong
absorption and emission in P~Cygni profiles. The absorption is weaker in the
model with lower mass-loss rate and less blue-shifted as a result of lower wind
terminal velocity. The carbon lines show P~Cygni profiles with slightly extended
absorption and show emission only for the highest mass-loss rates considered.
These lines become weaker close to the blue edge of the P~Cygni line profile due
to \ion{C}{ii} ionization. There are no such features in the observed spectra of
\hvezda. If the assumed abundances are correct, this would limit the \hvezda\
mass-loss rate to $\dot M\lesssim10^{-12}\,\msr$, below the value required to
explain the radio emission \citep{leto06}. On the other hand, the photospheric
components of C and Al lines show that these elements are likely underabundant
(similarly to carbon on the surface of $\sigma$~Ori~E, \citealt{mysigorie}). In
such a case the wind lines of these elements would not be present in the spectra
and the imposed wind limit should be higher.

Therefore, we concentrated on the wind lines of silicon and iron, the two
elements which are overabundant on the surface of \hvezda. The  1394\,\AA\ and
1403\,\AA\ lines of \ion{Si}{iv}, which might be relatively strong in the
spectra of stars with slightly higher effective temperature \citep{metuje}, do
not appear in the \hvezda\ spectra due to silicon recombination. Therefore, the
strongest limit to the mass-loss rate comes from the absence of \ion{Fe}{iii}
1895\,\AA\ wind line in Fig.~\ref{cuvir_met}, which gives $\dot
M\lesssim10^{-11}\,\msr$.

Spectra in Fig.~\ref{cuvir_met} are plotted for a particular phase
$\varphi\approx0.3$. In magnetic stars, the wind properties depend on the
tilt of the magnetic field \citep[e.g.,][]{brzdud}, therefore, the wind line
profiles show rotational variability \citep{bjorig,markomag,nasufu}. We tested
the existence of the wind also in other phases with similar conclusions as for
the phase $\varphi\approx0.3$. Consequently, we conclude that the presence of
the stellar wind in \hvezda\ can not be proven even from our spectra with the
highest quality. This corresponds to observation of very weak wind lines in
early-type magnetic stars \citep{oskibp}. These lines give the mass-loss rate on the order of $10^{-11}-10^{-10}\,\msr$, but for stars whose luminosity is nearly
two magnitudes higher. Given a strong dependence of mass-loss rate on
luminosity, the upper limit of the \hvezda\ mass-loss rate is not surprising.

There are additional possible wind indicators in the wavelength region
between the Lyman jump and Ly$\alpha$ line. The 1086\,\AA\ \ion{N}{ii} line is
predicted to show nice P~Cygni line profile, but it is located in the region
where the model flux is roughly one order of magnitude lower than the flux in
the 1300--1400\,\AA\ region. The lines 1176\,\AA\ \ion{C}{iii} and 1206\,\AA\
\ion{Si}{iii} are located in the wavelength region covered by the IUE
observations, but the observed flux is rather noisy. In any case, the observed
spectra in the region of 1206\,\AA\ \ion{Si}{iii} are consistent with the
$10^{-11}\,\msr$ limit of the mass-loss rate.

\hvezda\ has relatively strong magnetic field, which may affect the calculation
of the radiative force and wind line profiles \citep[e.g.,][]{shenmag}. The wind
lines originate in the magnetosphere with a complex structure leading to the
dependence of wind line profiles on phase and inclination
\citep[e.g.,][]{markomag}. The plasma effects become important in the wind at
low densities and the wind behaves as a multicomponent flow \citep{kkiii}.
Moreover, at low densities the shock cooling length is comparable to the radial
wind scale, possibly causing weak wind line profiles
\citep{cobecru,nlteiii,lucyjakomy}. We tested the inclusion of complete
photoionization cross-sections for \ion{Si}{ii} -- \ion{Si}{iii} (with
autoionization resonances, \citealt{maslo93}). This has not lead to a
significant increase of the radiative force, but there may still be some
elements missing in our list \citep{nlteiii} that are overabundant on \hvezda's
surface and may drive a wind. Moreover, wind inhomogeneities (clumping) affect
wind ionization state and wind line profiles \citep{chuchcar,clres1}. Although
\citet{oskibp} finds no significant differences in the spectra of clumped
and smooth stellar wind for their particular model of B star, there might still
be significant effects of clumping for other stellar parameters. Therefore,
there still might be possibilities to launch stellar wind in \hvezda\ that does
not show any signatures in UV spectra, but at the moment there is no theoretical
nor observational support for the existence of the stellar wind in \hvezda.

\section{Search for auroral lines}

The radio emission in \hvezda\ originates due to processes resembling
auroral activity in Earth and giant planets \citep{trigi11}. Therefore, we
searched for auroral lines in the UV spectra of \hvezda. Because the auroral
lines originate from the rigidly rotating magnetosphere, we expect these lines
to be broadened by at least $v_\text{rot}\sin i=145\,\text{km}\,\text{s}^{-1}$,
which corresponds to the projected equatorial rotational velocity
\citep{kocuvir}. Moreover, the auroral lines would likely show rotational
variability as a result of complex structure of the magnetosphere.

\begin{table}[t]
\caption{Lines searched in \hvezda\ spectra for auroral emission.}
\label{aurora}
\centering
\begin{tabular}{ll}
\hline
Ion & Lines [\AA]\\
\hline
\ion{He}{ii}  & 1640 \\
\ion{C}{i}    & 1330, 1561, 1657\\
\ion{C}{ii}   & 1335\\
\ion{O}{i}    & 1302, 1305, 1306, 1356, 1359\\
\ion{Si}{i}   & 1556 -- 1675\\
\ion{Si}{ii}  & 1305, 1309, 1527, 1533\\
\ion{P}{i}    & 1381 \\
\ion{P}{ii}   & 1533, 1536, 1542, 1543, 1544  \\
\ion{S}{i}    & 1283 -- 1475\\
\ion{Cl}{i}   & 1347, 1352\\
%\ion{Ti}{i}   & 1317, 1335, 1370, 1381, 1455 \\
\ion{Ti}{iii} & 1282 -- 1300\\
%\ion{V}{i}    & 1584 -- 1643\\
\ion{Fe}{ii}  & 1559 -- 1686\\
\ion{Co}{ii}  & 1448, 1456, 1466. 1473, 1509, 1548\\
\ion{Ni}{ii}  & 1317, 1335, 1370, 1411\\
\ion{Cu}{ii}  & 1359\\
\hline
\end{tabular}
\end{table}

The planetary auroral emission comprises molecular lines
\citep[e.g.,][]{sorac,gusta}, which are not expected in \hvezda\ due to its high
temperature. We searched for the auroral emission of neutral oxygen lines given
in Table~\ref{aurora}, which are detected in planetary magnetospheres
\citep[e.g.,][]{molina,sorac}. The lower level of all these lines corresponds to
the ground state. We find no auroral lines of neutral ions. This is
not surprising, as our wind models indicate that the magnetosphere is
dominated by singly and doubly ionized elements.

Therefore, we also searched for the auroral emission in lines whose
lower level corresponds to the ground state of singly and doubly ionized
elements (see Table~\ref{aurora}). We did not find any emission line profiles
that can be attributed to the circumstellar medium, either. However, this does
not necessary mean that there are no such lines present in the available
spectra. Given the complexity and variability of photospheric spectra, the
auroral lines may remain overlooked.

Some hot white dwarfs show ultra-high excitation absorption lines in their
spectra \citep[e.g.,][]{uhl,dreiuhe,uhlprom}. These lines possibly originate in
shock-heated magnetospheres of white dwarfs. Such a process can also appear in
magnetic main-sequence stars. We searched for these lines in the optical spectra
of \hvezda, but we have not found any.

\section{Lines of interstellar medium}

\begin{figure}[tp]
\centering \resizebox{\hsize}{!}{\includegraphics{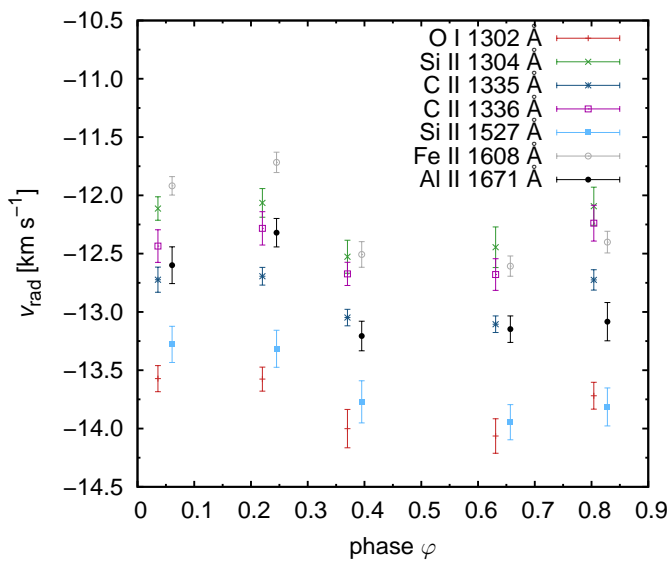}}
\caption{Variation of the radial velocity shift of the interstellar lines with
phase. Derived from $\delta\lambda$ parameter of the fit \eqref{ism} using the
Doppler law.}
\label{analyzacar_e}
\end{figure}

HST/STIS spectra of \hvezda\ show several interstellar lines, which are missing
in Figs.~\ref{hst1380} and \ref{hst1598} due to the application of median
filter. We studied the variability of interstellar lines in the original spectra
downloaded from MAST archive. For this purpose we fitted the individual
interstellar lines by 
\begin{equation}
\label{ism}
F(\lambda)=(F_0+F_1\lambda)
\exp\szav{-s\exp\hzav{-\zav{\frac{\lambda-\lambda_0-\delta\lambda}{d}}^2}},
\end{equation}
which is a solution of the radiative transfer equation in a uniform medium with
Gaussian line absorption. Parameters $F_0$ and $F_1$ account for the stellar
continuum, $\lambda_0$ is the laboratory wavelength of a given line, and $s$,
$\delta\lambda$, and $d$ are parameters of the fit.

We identify individual interstellar lines using the list provided by
\citet{cox}. We fitted them by line profile function Eq.~\eqref{ism} and
searched for the phase variability of the line parameters. We find phase
variability of the line shift $\delta\lambda$ (see Fig.~\ref{analyzacar_e}),
which is correlated with line width parameter $d$ in both STIS bands.

\begin{table}[t]
\caption{Parameters of the fit Eq.~\eqref{ism} of individual interstellar lines.}
\label{ismtab}
\centering
\begin{tabular}{ccc}
\hline
Line & $\delta\lambda$ [m\AA] & $d$ [m\AA]\\
\hline
\ion{O}{i}   1302\,\AA & $-59.9\pm0.8$ & $23.8\pm1.4$\\
\ion{Si}{ii} 1304\,\AA & $-53.3\pm0.8$ & $19.2\pm1.1$\\
\ion{C}{ii}  1335\,\AA & $-57.3\pm0.7$ & $28.0\pm0.7$\\
\ion{C}{ii}  1336\,\AA & $-55.5\pm0.7$ & $13.6\pm0.4$\\
\ion{Si}{ii} 1527\,\AA & $-69.4\pm1.2$ & $24.4\pm0.7$\\
\ion{Fe}{ii} 1608\,\AA & $-65.6\pm1.6$ & $15.4\pm0.5$\\
\ion{Al}{ii} 1671\,\AA & $-71.7\pm1.7$ & $22.6\pm0.5$\\
\hline
\end{tabular}
\end{table}

It is tempting to attribute these variations to circumstellar clouds that
amplify the interstellar absorption. The centrifugally supported circumstellar
clouds can exist above the radius where the centrifugal force overcomes gravity
$r=\hzav{GMP^2/\zav{4\pi^2}}^{1/3}$ \citep{towo}, which for the stellar
parameters given in Table~\ref{hvezda} gives $r=1.9\,R_*$. This is at odds with
estimated temperature of the clouds. The \ion{Si}{ii} 1533\,\AA\ line with
lower-level excitation energy 0.036\,eV is not present in the spectrum, while
\ion{C}{ii} 1336\,\AA\ line with lower-level excitation energy 0.008\,eV is
clearly visible. This implies the temperature of absorbing medium on the order
of 10~K, which cannot be achieved in a close proximity of the star. Moreover,
the expected rotational line broadening due to rigidly rotating centrifugal
magnetosphere, which is at least equal to the surface rotational velocity
$v_\text{rot}\sin i=145\,\text{km}\,\text{s}^{-1}$ \citep{kocuvir}, is an order
of magnitude higher than the observed broadening of interstellar lines. The
difference between the shift estimated from the spectra centered at 1380\,\AA\
and spectra centered at 1598\,\AA\ around the phase $\varphi\approx0.8$
(Fig.~\ref{analyzacar_e}) is comparable to the uncertainty of $\delta\lambda$.
Finally, the unsaturated \ion{C}{ii} 1336\,\AA\ line shows the same variability
as saturated lines implying comparable absorption due to both the interstellar medium
and to the putative circumstellar cloud. Consequently, we conclude that the
line variability in Fig.~\ref{analyzacar_e} does not originate in the
centrifugal magnetosphere and that it is likely generated further out from the
star.

We also tested the possibility that the interstellar line variability is a
result of a constant circumstellar absorption acting on variable flux. For this
purpose we used simulated spectra attenuated by circumstellar line absorption
and fitted the resulting spectra by Eq.~\eqref{ism}. We detected only order of
magnitude smaller variations than those derived from observations. Consequently,
the observed variations of the interstellar medium lines are most likely only of
instrumental origin. Although the STIS spectrograph provides subpixel wavelength
stability (i.e., leading to lower changes than that seen in
Fig.~\ref{analyzacar_e}, \citealt{stis}), subsequent processing of variable
spectra likely leads to larger differences. We therefore simply averaged the
radial velocities inferred from Table~\ref{ismtab} to get the mean radial
velocity of the interstellar medium of $-12.9\pm0.4\,\text{km}\,\text{s}^{-1}$.
Because \hvezda\ is located close to the rim of the Spica Nebula, which results
from the interaction of the Local Bubble with another interstellar bubble
\citep{choi}, the interstellar lines in \hvezda\ spectra may be connected
with these structures.

\section{Discussion: Tracing the wind}

The upper limit for the wind mass-loss rate $10^{-12}\,M_\odot\,\text{yr}^{-1}$
is in tension with the mass-loss rate on the order of
$10^{-12}\,M_\odot\,\text{yr}^{-1}$ required to explain the \hvezda\ radio
emission \citep{leto06}. This estimate was derived assuming the balance between
the ram pressure of the wind and the plasma pressure in the inner magnetosphere.
On the other hand, leaving this requirement and assuming instead that the
magnetospheric plasma is in hydrostatic equilibrium \citep{towo} allows for much
lower wind mass-loss rate required to feed the magnetosphere. Within such a
model, even the pure metallic wind predicted by \citet{babelb} might be
strong enough to power the radio emission.

In contrast, even relatively strong wind is consistent with the angular momentum
loss via the magnetized stellar wind. This mechanism affects the rotational
velocities of early-B stars \citep{pirat} and should be present also in late-B
stars with winds. Assuming $\dot M=10^{-12}\,\msr$, a typical wind terminal
velocity $v_\infty=1000\,\text{km}\,\text{s}^{-1}$, dipolar magnetic field
strength of \hvezda\ $B_\text{p}=3\,\text{kG}$ \citep{kocuvir}, and angular
momentum constant $k=0.08$ as derived from MESA models \citep[see
\citealt{krtvar} for details]{mesa1,mesa2} we derive from Eq.~(25) of
\citet{brzdud} the spin-down time $1.4\times10^8\,\text{yr}$, which is
consistent with an estimated age $\leq100\,\text{Myr}$ of \hvezda\
\citep{kocuvir}.

The stellar wind also likely powers X-ray emission of \hvezda\
\citep{takyjsemnasel}. Assuming full conversion of wind kinetic energy to
X-rays, then with wind terminal velocity
$v_\infty=1000\,\text{km}\,\text{s}^{-1}$, a mass-loss rate
$10^{-13}\,{M}_\odot\,\text{yr}^{-1}$ is required to power the X-ray luminosity
$L_\text{X} = 3 \times 10^{28}\,\text{erg}\,\text{s}^{-1}$ derived from the
observations. This mass-loss rate is below the upper limit derived here.

It is unclear what drives the supposed torsional oscillations in \hvezda. The
properties of the corresponding energy source can be estimated from the total
energy of these oscillations, which is given by the kinetic energy density of
oscillatory motion integrated over the whole stellar volume,
\begin{equation}
E_\text{to}=4\pi\int_0^{R_*}r^2\frac{1}{2} \rho\delta \varv^2 \,\de r,
\end{equation}
where $\delta \varv$ is the velocity amplitude of the oscillations, which is
connected with the amplitude of the angular frequency by $\delta \varv
=r\delta\Omega$. The surface value of the amplitude of the angular frequency
$\delta\Omega_*$ can be derived from the harmonic fit of the period
variability $P(t)=P_0+A\,\sin\hzav{2\,\pi\zav{t-\mathit{\Theta_0}}/
{\mathit{\Pi}}}$ \citep{krtvar} as $\delta\Omega_*\approx2\pi A/P_0^2$.
Here $A$ and $\mathit{\Pi}$ are the amplitude and length of the cycle of
rotational period evolution. From
this, the total energy of the torsional oscillations is
\begin{equation}
\label{eto}
E_\text{to}=8\pi^3\frac{A^2}{P_0^4}\int_0^{R_*}r^4\rho
\zav{\frac{\delta\Omega}{\delta\Omega_*}}^2\,\de r.
\end{equation}
The energy can be conveniently rewritten using the angular momentum constant
$k$, and assuming that $\delta\Omega$ does not depend on radius one can get an
estimate $E_\text{to}=2\pi^2kMR^2A^2/P_0^4$.

To derive a more precise estimate of the total energy of the torsional
oscillations, we used the stellar density distribution from the discussed MESA
stellar evolutionary model. We assumed that the amplitude of $\delta\Omega$
is proportional to the amplitude of $u_1$ variable for time $t=10\,\mathit{\Pi}$
of torsional oscillation simulations \citep[Fig.~1]{krtvar}. The selection of a
particular time $t$ does not have a significant effect on estimated value of
$E_\text{to}$. We scaled the numerical solution for $u_1$ using the
observed variations of rotational period at the stellar surface. These
assumptions yield $E_\text{to}= 5\times10^{37}\,\text{erg}$, which is just about
$5\times10^{-7}$ of the total energy of X-rays emitted during the whole \hvezda\
lifetime (assuming constant $L_\text{X}$ and an upper limit of \hvezda\ age,
100\,Myr, \citealt{kocuvir}). Consequently, stellar wind, which presumably
powers the X-ray emission, provides a convenient energy source also for the
torsional oscillations.

The continuous wind leakage from the magnetosphere is likely not the mechanism
that drives the oscillations, but a sudden energy release during the
reconnection event that opens up the magnetosphere \citep{towo} might be more
likely. Such events are probably very rare, because the corresponding breakout
time is about $10^5\,\text{yr}$ as estimated from Eq.~(A8) of \citet{towo} for
\hvezda\ parameters.

\section{Conclusions}

We analyzed the HST/STIS spectra of the enigmatic chemically peculiar star
\hvezda. This is the first discovered main-sequence star showing coherent
directive radio emission \citep{trigilio}. The existence of such radio emission
requires wind with mass-loss rate on the order of
$10^{-12}\,{M}_\odot\,\text{yr}^{-1}$ as a source of free electrons
\citep{leto06}. However, the effective temperature of \hvezda\ is below the
limit of homogeneous winds $T_\text{eff}\approx15\,000\,$K \citep{metuje}. Our
analysis of UV spectra of the star has not revealed any wind signatures and puts
the upper limit of \hvezda\ mass-loss rate at about
$10^{-12}\,{M}_\odot\,\text{yr}^{-1}$. Although the physics of low density winds
is complex and there are reasons for even stronger winds to remain hidden from
UV analysis, we argue that even weaker inhomogeneous (metallic) wind of
\citet{babelb} may be a reasonable source of free electrons to power the radio
and X-ray emission of \hvezda.

We searched for other signatures of magnetospheric plasma including auroral
emission lines and deep absorption lines. However, we have not found any
signatures that can be attributed to magnetospheric plasma.

\hvezda\ shows long-term rotational period variations \citep{pyperper,mikvar}.
We analyzed these variations using available UV observations supplemented with
our own and archival optical photometric data. We concluded that the data from
optical and UV domains show the same long-term evolution of the rotational
period. This supports the picture that the UV and optical variations are
related and that the structures leading to these variations are rigidly
confined to the stellar surface.

We modeled the UV flux variability assuming its origin in the surface abundance
spots, in which the flux redistribution due to bound-bound and bound-free
processes takes place. We show that the updated abundance maps of
\citet{kocuvir} provide improved flux distributions that agree better with
observations mainly around $1600\,\AA$.

\begin{acknowledgements}
This research was supported by grants GA\,\v{C}R  16-01116S
and in final phases by GA\,\v{C}R 18-05665S. GWH acknowledges
long-term support from NASA, NSF, and the Tennessee State University Center of
Excellence, as well as NASA grant HST-GO-14737.002-A. OK acknowledges support by
the Knut and Alice Wallenberg Foundation, the Swedish Research Council, and the
Swedish National Space Board. AP acknowledges the support from the National
Science Centre grant no. 2016/21/B/ST9/01126. This research was partly based on
the IUE data derived from the INES database using the SPLAT package.
\end{acknowledgements}

\end{document}